\documentclass[prb,aps,showpacs,floats,twocolumn]{revtex4}
\usepackage{graphicx}

\begin{document}
\title{Correlation-driven phase transition in a Chern insulator}
\author{Hong-Son Nguyen$^1$ and Minh-Tien Tran$^2$}
\affiliation{$^1$Department of Occupational Safety and Health, Trade Union University, 169 Tay Son, Hanoi, Vietnam \\
$^2$Institute of Physics, Vietnam Academy of Science and Technology, 10 Dao Tan, Hanoi, Vietnam}

\begin{abstract}
The phase transition driven by electron correlations in a Chern insulator is investigated within the dynamical mean-field theory. The Chern insulator is described by the Haldane model and the electron correlations are incorporated by introducing the short-range interaction between the itinerant electrons and localized fermions. In the preservation of the inversion symmetry, the electron correlations drive the system from the Chern insulator to a renormalized pseudogap metal, and then to the topologically trivial Mott insulator. When the inversion symmetry is broken, a charge ordering and a nontrivial Chern topological invariant coexist.

\end{abstract}


\maketitle

\section{Introduction}
Recently, the theoretical prediction and experimental discovery of the topological insulator with the time-reversal symmetry, the $Z_2$ topological insulator, has attracted a lot of attention in condensed matter physics.\cite{Bernevig,Konig,Hassan,Zhangrv} The topological phase is beyond the Landau phase concept and characterized by nontrivial topological invariants that lead the spin Hall conductance to be nonzero quantized. The topological insulators are gapped in the bulk but have gapless edge state protected by the time-reversal symmetry. The concept of the topological insulator has also been expanded into the lattice symmetry protected topological phases which exhibit topological invariants as long as the symmetries are not broken.\cite{Fang}

Before the discovery of the $Z_2$ topological insulator, the quantum Hall states were the well-known realization of the topological phases. They exhibit the quantized Hall conductance. However, the quantum Hall states do not have the time-reversal symmetry as well as the lattice translational symmetry. In addition, the quantum Hall states require an external magnetic field that forms the Landau levels. Haldane theoretically proposed a lattice model that also exhibits an integer Hall conductance, but in the absence of external magnetic fields and the Landau levels.\cite{Haldane} It turns out that the integer Hall conductance is just the Chern topological invariant.\cite{TKNN} Such topological states are usually called the Chern insulator. In the Chern insulator, the time-reversal symmetry is broken.
There were several proposals to realize the Chern topological phase following the Haldane idea. They include the proposals of the realization of the Haldane model by loading ultracold atoms into optical lattices.\cite{Shao,Stanescu1,Stanescu2} Recently, the Chern insulator was discovered in thin films of Cr-doped (Bi,Sb)$_2$Te$_3$,\cite{Chang} following the theoretical prediction.\cite{Yu} In the thin films, the ferromagnetic order breaks the time-reversal symmetry, and this effect is essentially described by electron correlations.

The theory of both the $Z_2$ and Chern topological insulators is based on the energy band theory, where electron correlations are weak. The effect of electron correlations on the topological phases has also attracted much attention.\cite{Assaad} One may expect that weak electron correlations do not change the topological properties of the topological insulators. However, strong electron correlations can give rise to qualitative changes of their properties. Most studies have focused on the correlation effects in the $Z_2$ topological insulators. The electron correlations can drive the topological insulating phase to the Mott or Slater insulating phases. The Slater transition is accompanied by a long-range order, whereas in the Mott transition spontaneous symmetry breaking is absent.\cite{Mott} In particular, when the local Coulomb interaction is included in the Kane-Mele model of the $Z_2$ topological insulator,\cite{KaneMele} it drives the band topological insulating state into a spin-density-wave state, where the time-reversal symmetry is broken.\cite{Hur1,Hur2,Zheng,Meng,YuLi} Depending on the model parameters, the phase transition may go through a spin-liquid phase.\cite{Hur2,Zheng,Meng,YuLi} However,  there are still debates over the existence of the spin-liquid state.\cite{Assaad}

In contrast to the $Z_2$ topological insulators, the effects of electron correlations on the Chern insulators have received less attention. Originally, the Haldane model, which describes a Chern insulator, was a spinless fermion model. Most studies have introduced nonlocal nearest-neighbor Coulomb interactions into the Haldane model and investigated their effects on the topological properties.\cite{Raghu,Galitski1,Galitski2} The nearest-neighbor interactions drive the system into a charge-ordered insulating phase. The charge-ordered phase is topologically trivial, and the phase transition is of first order.\cite{Raghu,Galitski1} While the time-reversal symmetry is broken in both the Chern-insulating and the charge-ordered phases, the inversion symmetry, which is preserved in the Chern-insulating phase, is lost in the charge-ordered phase. This situation is similar to the Slater transition in the $Z_2$ topological insulators, where the transition is accompanied by symmetry breaking. However,
it is still not clear what happens to the Mott transition, which is not accompanied by any symmetry breaking, in the topological insulators.

In the present paper we study the effects of electron correlations in a Chern insulator that is described by the Haldane model. In contrast to the previous studies,\cite{Raghu,Galitski1,Galitski2} we incorporate a local interaction into the Haldane model. The local interaction is the short-range Coulomb interaction between the itinerant electrons and additional localized spinless fermions. This interaction is essentially the type of the local interaction in the Falicov-Kimball model.\cite{Falicov} The Haldane model with the Falicov-Kimball interaction can also be considered as an asymmetric version of the Kane-Mele-Hubbard model, where electrons with a fixed spin component are frozen. The time reversal symmetry is preserved in the Kane-Mele-Hubbard model, but it is broken in the Haldane-Falicov-Kimball model. It is well known that the Falicov-Kimball model describes well the correlation-induced  phase transitions with and without inversion-symmetry breaking.\cite{Falicov,Kennedy,Gruber1,Gruber2} One may expect that with the local Coulomb interaction the introduced model is suitable for studying the topological phase transitions due to electron correlations with and without inversion-symmetry breaking in Chern insulators. We employ the dynamical mean-field theory (DMFT) to study the phase transition. The DMFT was widely and successfully used to study electron correlations.\cite{Metzner,GKKR} In particular, the Falicov-Kimball model was intensively studied within the DMFT.\cite{Mielsch1,Mielsch2,Mielsch3,Freericks}  We find that when the inversion symmetry is preserved, electron correlations drive the system from the topological Chern insulator to a pseudogap metallic state, and then to the topologically trivial Mott insulator. The metallic state is non-Fermi-liquid with the renormalized mass and velocity of Weyl fermions. It always exists between the Chern and the Mott insulators. When the inversion symmetry is broken, electron correlations form a charge-ordering gap, and at the same time they drive the system from the topological charge-ordered state to the topologically trivial charge-ordered state. Both the topological invariant and the long-range charge order coexist in the topological charge-ordered state.

The present paper is organized as follows. In Sec. II we introduce the Falicov-Kimball type of the local interaction to
the Haldane model. In this section we also present the DMFT for the proposed model. The Mott transition is presented in Sec. III, while the charge-ordering transition is presented in Sec. IV. Finally, in Sec. V we present the conclusion.

\section{The Haldane model with local interaction}
In this section we introduce a local interaction into the Haldane model. The Haldane model describes the hopping of noninteraction spinless electrons with zero net magnetic flux in a honeycomb lattice.\cite{Haldane} In addition to the itinerant electrons, we introduce localized spinless fermions. Without interaction, the localized fermions do not affect the topological properties of the itinerant electrons. A local interaction which is the Coulomb interaction between the spinless itinerant electrons and localized fermions is also introduced. The total model is described by the Hamiltonian
\begin{eqnarray}
H &=&-t\sum\limits_{<i,j>}c_{i}^{\dagger }c_{j}+ i t_{2}\sum\limits_{\ll i,j\gg
} \nu_{ij}c_{i}^{\dagger }c_{j}+{\rm H.c.}  \nonumber \\
&& + E_f \sum_{i} f^{\dagger}_{i} f_{i} + U \sum_{i} c^{\dagger}_i c_i f^{\dagger}_i f_i,
\label{hfk}
\end{eqnarray}
where $c_i$ ($f_i$) are the annihilation operators for itinerant (localized) spinless fermions at site $i$ of a honeycomb lattice. $t$ is the parameter of the nearest-neighbor hoppings, whereas $t_2$ is the one of the next-nearest-neighbor hoppings. $\nu_{ij}=\pm 1$ for the clockwise (anticlockwise) next-nearest-neighbor hoppings. $E_f$ is the energy level of the localized fermions. $U$ is the strength of the local interaction between the itinerant and localized fermions. When $U=0$, the Hamiltonian in Eq. (\ref{hfk}) describes the Haldane model with the Peierls phase $\pi/2$, and zero staggered energy shift of the two  sublattices of the honeycomb lattice.\cite{Haldane} The ground state is essentially topological with the Chern number $C=\pm 1$, depending on the sign of $t_2$.\cite{Haldane} When $t_2=0$, the Hamiltonian in Eq.  (\ref{hfk}) is the standard Falicov-Kimball model on a bipartite lattice.\cite{Falicov} It is well known that the Falicov-Kimball model on a bipartite lattice exhibits the metal-insulator transition for the homogeneous states, where the inversion symmetry is preserved, and charge ordering where the inversion symmetry is broken.\cite{Falicov,Kennedy,Gruber1,Gruber2} In the other limit, when $t=0$, the proposed model is equivalent to two independent triangular lattices of the Falicov-Kimball model. The triangular lattice is geometrically frustrated; thus, the regular charge ordering as the one in the square lattice does not occur at low temperatures.\cite{Gruber3,Cenca1,Cenca2,Yadav} Instead, striped or bounded phases are formed.\cite{Gruber3,Cenca1,Cenca2,Yadav} However, one might expect that the next-nearest-neighbor hopping $t_2$ is always smaller than the nearest-neighbor hopping $t$, and for small values of $t_2$ the frustrations of the individual sublattices do not break the bipartite structure of the whole honeycomb lattice. In the rest of paper we only consider the regime  $t_2<t$.
The introduced model in Eq. (\ref{hfk}) can also be considered as an asymmetric version of the Kane-Mele-Hubbard model,\cite{Assaad} where the electrons with a fixed spin component are frozen. The asymmetry can occur as a consequence of the extreme mass imbalance of the two spin components. Electrons with a fixed spin component are extremely heavy, and come to be localized. Due to the asymmetry the time-reversal symmetry is explicitly broken. However, the Hall conductance is still quantized in the noninteraction case.   Incorporating the Haldane idea of the topological phase into the Falicov-Kimball model, it gives rise to a possibility of  studying the correlation-driven topological phase transition with and without inversion-symmetry breaking.

We use the DMFT to investigate the correlation-driven phase transition in the introduced model in Eq. (\ref{hfk}). Within the DMFT, the self-energy only depends  on frequency. It is exact in infinite dimensions. However, in two-dimensional systems it is an approximation. The approximation neglects nonlocal correlations, but keeps the local dynamical correlations. The DMFT applied to the Hubbard model in the honeycomb lattice overestimates the critical point of the Mott transition.\cite{Sorella,Tran,WuLiu} The Falicov-Kimball model was also studied within the DMFT.\cite{Mielsch1,Mielsch2,Mielsch3,Freericks} Since the honeycomb lattice is bipartite, for convenience, we divide the lattice into two penetrating sublattices $A$ and $B$, like the DMFT of the Falicov-Kimball model on hypercubic lattices.\cite{Mielsch1,Mielsch2,Mielsch3,Freericks,Zlati,Krish,Matveev} In the momentum space, the hopping part of the Hamiltonian in Eq. (\ref{hfk}) written in the two sublattice indexes is a $2\times 2$ matrix
\begin{equation}
\widehat{h}_{0}(\mathbf{k})=\left(
\begin{array}{cc}
t_{2}f_{2}(\mathbf{k}) & -tf_{1}(\mathbf{k}) \\
-tf_{1}^{\ast }(\mathbf{k}) & -t_{2}f_{2}(\mathbf{k})
\end{array}%
\right),
\end{equation}
where
\begin{eqnarray*}
f_{1}(\mathbf{k}) &=& e^{i\frac{1}{2}k_{x}}\cos \big(\frac{\sqrt{3}}{2}k_{y}\big)+e^{-ik_{x}} , \\
f_{2}(\mathbf{k}) &=& 2\sin \big(\frac{3}{2}k_{x}+\frac{\sqrt{3}}{2}k_{y}\big)-2\sin \big(\frac{3}{2}k_{x}-
\frac{\sqrt{3}}{2}k_{y}\big) \\
&& -2\sin (\sqrt{3}k_{y}).
\end{eqnarray*}
Within the DMFT, the Green's function of the itinerant electrons can be written via the Dyson equation
\begin{eqnarray}
\hat{G}(\mathbf{k},z) = \big[ z+\mu-\hat{h}_{0}(\mathbf{k}) -\hat{\Sigma}(z)
\big]^{-1} ,
\end{eqnarray}
where $\mu$ is the chemical potential, and
$$\hat{\Sigma}(z) =\left(\begin{array}{cc}
\Sigma_{A}(z) & 0 \\
0 & \Sigma_{B}(z)
\end{array}
\right)
$$
is the self-energy. The self-energy is determined from a chosen single site embedded in an effective mean-field medium which represents the interactions of all other sites among themselves and with the chosen site. The action of the effective single site is
\begin{eqnarray}
S_{\alpha} &=&\int d\tau d\tau^{\prime }c^{\dagger }_{\alpha}(\tau )\mathcal{G}
_{\alpha}^{-1}(\tau -\tau ^{\prime })c_{\alpha}(\tau^{\prime} )  \nonumber \\
&& +(E_{f}-\mu )f^{\dagger}_{\alpha}f_{\alpha}
+Uc^{\dagger }_{\alpha}c_{\alpha} f^{\dagger }_{\alpha}f_{\alpha},
\label{action}
\end{eqnarray}
where $\alpha=A,B$ is the sublattice index. $\mathcal{G}_{\alpha}(\tau)$ is the Green's function that represents the effective medium.
It can be written via $\mathcal{G}^{-1}_{\alpha}(z)=z+\mu-\lambda_{\alpha}(z)$, where $\lambda_{\alpha}(z)$ can be considered as a dynamical Weiss mean field. The self-energy of the effective single site satisfies the Dyson equation
\begin{equation}
G^{-1}_{\alpha}(z)=z+\mu-\lambda_{\alpha}(z)-\Sigma_{\alpha}(z) . \label{gl}
\end{equation}
The self-consistent condition requires that the Green's function obtained from the effective site and the local Green's function of the lattice are identical; i.e.,
\begin{equation}
G_{\alpha}(z) = \frac{1}{N} \sum_{\mathbf{k}} \hat{G}_{\alpha\alpha}(\mathbf{k},z) , \label{sc}
\end{equation}
where $N$ is the number of sublattice sites.

Effective action in Eq. (\ref{action}) can analytically be solved, since $f^{\dagger}f$ is a good quantum number. We obtain the partition function
\begin{eqnarray}
\lefteqn{
Z_{\alpha} = {\rm Tr}_{f} \int \mathcal{D}[c^{\dagger },c]e^{-S_{\alpha}[c^{\dagger },c,f^{\dagger},f]} }  \nonumber \\
&&= 2\exp \left[ \sum\limits_{n}\ln \left( \frac{i\omega _{n}+\mu -\lambda
_{\alpha}(i\omega _{n})}{i\omega _{n}}\right) \right]
 +e^{-\beta (E_{f}-\mu)} \nonumber \\
 && 2 \exp\bigg[\sum\limits_{n}\ln \left( \frac{i\omega _{n}+\mu -\lambda
_{\alpha}(i\omega _{n})-U}{i\omega _{n}}\right) \bigg] , \label{part}
\end{eqnarray}
where $\omega_n=(2n+1) \pi T$ is the Matsubara frequency. From the partition function in Eq. (\ref{part}) we obtain the Green's function
\begin{eqnarray}
G_{\alpha}(z)
=\frac{1-n_{f\alpha}}{z+\mu-\lambda_{\alpha}(z)}+\frac{n_{f\alpha}}{z+\mu-\lambda_{\alpha}(z)-U} , \label{gli}
\end{eqnarray}
where
\begin{eqnarray}
n_{f\alpha} &=& \frac{1}{1+\exp [\beta (\widetilde{E}_{f\alpha}-\mu)]} , \label{nf} \\
\widetilde{E}_{f\alpha} &=& E_{f}  +T\sum\limits_{n}\ln \left( \frac{i\omega
_{n}+\mu -\lambda _{\alpha}(i\omega _{n})}{i\omega _{n}+\mu -\lambda _{\alpha}(i\omega
_{n})-U}\right) . \nonumber
\end{eqnarray}
One can check that $n_{f\alpha}$ is exactly the average number of localized fermions in sublattice $\alpha$.  When $n_{fA}=n_{fB}$ the two sublattices are equivalent, and the inversion symmetry is preserved.
When $n_{fA} \neq n_{fB}$, a charge ordering occurs and the inversion symmetry is broken. From Eq. (\ref{nf}) one can see that at zero temperature $n_{f\alpha}$ can accept only three values: $0$, $1/2$, and $1$. For half filling two nonequivalent possibilities can occur: (i) $n_{fA}=n_{fB}=1/2$; (ii) $n_{fA}=1$, $n_{fB}=0$ (or $n_{fA}=0$, $n_{fB}=1$). In the first case, the renormalized energy level $\widetilde{E}_{fa}$ of the localized electrons is pinned to the Fermi level,\cite{Si} while in the second case it is lower (or higher) the Fermi level. We numerically solve the self-consistent equations of the DMFT  in both real and imaginary frequencies by iterations. At zero temperature the imaginary frequencies are understood as the Matsubara frequencies with a fictitious temperature.

\section{Mott transition}

\begin{figure}[t]
\includegraphics[width=0.45\textwidth]{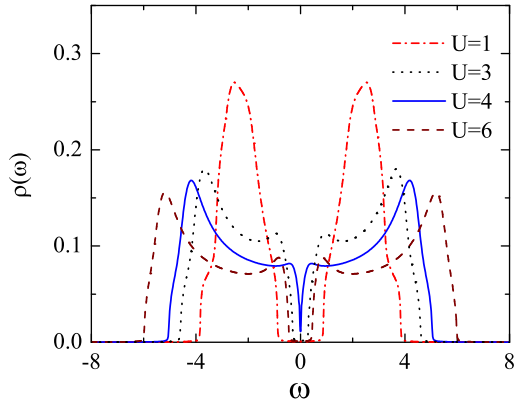}
\caption{(Color online) The DOS for various values of $U$ ($t=1$, $t_2=0.5$).} \label{fig1}
\end{figure}

\begin{figure}[t]
\includegraphics[width=0.45\textwidth]{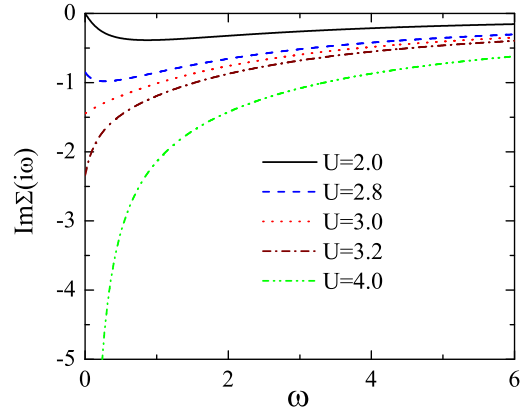}
\caption{(Color online) The imaginary part of the self energy $\rm{Im} \Sigma(i\omega)$ for various values of $U$ ($t=1$, $t_2=0.1$).} \label{fig2}
\end{figure}

We consider the half-filling case. It turns out $\mu=U/2$. When $n_{fA}=n_{fB}=1/2$ the Mott transition occurs. In Fig. \ref{fig1} we plot the density of states (DOS) $\rho(\omega)=-{\rm Im}G_{\alpha}(\omega+i0^{+})/\pi$ for various values of $U$. It shows that with increasing $U$ the DOS first exhibits a band gap, then the gap closes, and finally it opens again. These behaviors indicate the phase transition from an insulating phase to a metallic phase, and then to an insulating phase again. Since when $U=0$ the system is a Chern insulator, and the region of small values of $U$ adiabatically connects with $U=0$, one can expect that the first insulating phase is a Chern insulator. In this region, the DOS clearly shows two separated subbands. The subband separation increases with $U$. However, the actual gap, where the DOS vanishes, decreases with increasing $U$. While electron correlations try to separate the two subbands, they actually also reduce the gap of the Chern insulator. At critical value $U_{c1}$ the gap closes, and the system becomes metallic. We verify the topology of the first insulating phase by calculating directly the Chern number. In the interaction case the Chern number $C$ can be calculated via the Green's function at zero frequency\cite{Volovik,Gurarie,Wang}
\begin{eqnarray}
C = \frac{1}{2\pi} \int d^2 k \mathcal{F}_{xy} , \label{chern}
\end{eqnarray}
where $\mathcal{F}_{ij}=\partial_{i} \mathcal{A}_{j} - \partial_{j} \mathcal{A}_{i}$,
$\mathcal{A}_{i}=-i \sum_{\nu}^{'} \langle \mathbf{k} \nu | \partial_{k_{i}} |\mathbf{k} \nu \rangle$, and the sum is taken over all orthonormalized eigenstates $|\mathbf{k} \nu \rangle$ of matrix $\hat{G}^{-1}(\mathbf{k},i0)$ with positive eigenvalues. In numerical calculations we use the efficient method of discretization of the Brillouin zone to calculate the Chern number in Eq. (\ref{chern}).\cite{Fukui} In particular, we use a fine mesh of $128\times 128$ points for the elementary cell of the reciprocal lattice. Indeed, we obtain $C=1$ when $U<U_{c1}$. The first insulator-metal transition is also the topological phase transition. The gap closes at the same point where the Chern number stops to be quantized. The metallic phase, where the DOS is finite at the Fermi energy, occurs for $U$ ran from $U_{c1}$ up to $U_{c2}$. At $U_{c2}$ the gap opens again.
Although the actual gap closes in the metallic phase, the DOS clearly exhibits a pseudogap; in particular, a narrow dip exists around $\omega=0$.
The last insulating phase occurs for strong correlations $U>U_{c2}$, and it is naturally a Mott insulator. We can distinguish the phases by considering the behavior of the self-energy at low frequencies. In Fig. \ref{fig2} we plot the imaginary part of the self-energy $\rm{Im} \Sigma(i\omega)$ for various values of $U$. In the Chern-insulator phase the imaginary part of the self-energy vanishes, $\rm{Im} \Sigma(i0)=0$, which leads to the DOS also vanishing at $\omega=0$. In the metallic phase  $\rm{Im} \Sigma(i0)$ is finite. This indicates that the metallic phase is a non-Fermi-liquid. The slope of $\rm{Im} \Sigma(i\omega)$ at $\omega \rightarrow 0$ is identical to the slope of $\rm{Re} \Sigma(\omega+i0^{+})$ at $\omega \rightarrow 0$. Then we can use the slope of
$\rm{Im} \Sigma(i\omega)$ at $\omega \rightarrow 0$ to determine the renormalized factor
\begin{figure}[b]
\vspace{0.1cm}
\includegraphics[width=0.45\textwidth]{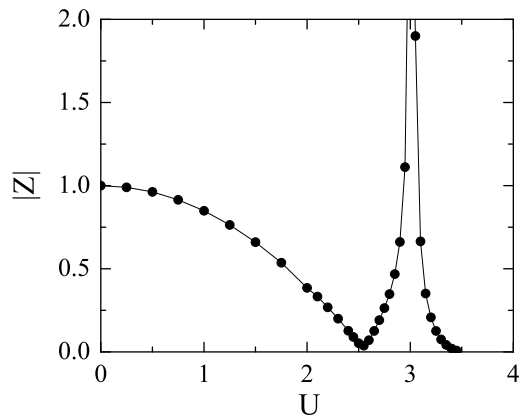}
\caption{The absolute value of the renormalized factor $|Z|$ as a function of $U$ ($t=1$, $t_2=0.1$).} \label{fig2a}
\end{figure}
\begin{equation}
Z = \bigg[ 1- \frac{\partial \rm{Re} \Sigma(\omega+i0^{+})}{\partial \omega}
\bigg]^{-1}_{\omega=0} .
\end{equation}
In Fig. \ref{fig2a} we plot the absolute value of the renormalized factor $|Z|$ as a function of $U$. It shows that in the metallic phase the renormalized factor diverges when $\partial \rm{Im}\Sigma(i\omega)/\partial \omega =1$ at $\omega=0$. Away from the divergence point, $|Z|$ decreases, and it vanishes at both $U_{c1}$ and $U_{c2}$. However, when $U<U_{c1}$, the renormalized factor remains finite. This indicates that the Chern-insulator phase remains the band insulator like the noninteraction case. The renormalized factor always vanishes when $U>U_{c2}$.
 This mimics the Brinkman-Rice scenario of the Mott transition.\cite{Brinkman}
The dip feature of the DOS around zero energy in the metallic phase is due to the special property of the honeycomb lattice. When $t_2=0$ and $U=0$, at the corners of the first Brillouin zone electrons become the Weyl fermions with linear dispersion $\varepsilon_{k} = v k$, which leads to the DOS linearly vanishing at zero energy.\cite{Neto} With finite hopping $t_2$, electrons acquire a mass at the corners of the first Brillouin zone. They also get a finite lifetime when electron correlations are in effect. The electron correlations also renormalize both the mass and the velocity of electrons at the corners of the the first Brillouin zone through the renormalized factor $Z$. In consequence, the DOS exhibits the dip around the Fermi energy, but its value is always finite. In the Mott-insulator phase
the imaginary part of the self-energy  $\rm{Im} \Sigma(i\omega)$ diverges like $1/\omega$. Due to this divergence we cannot use Eq. (\ref{chern}) to calculate the Chern number in the Mott-insulator phase. However, due to the local feature of the self-energy, the Chern number can be determined through a frequency domain winding number (FDWN);\cite{Xie} i.e.,
\begin{eqnarray}
C &=& \gamma C_{0} , \nonumber \\
C_{0} &=& \int \frac{d^2 k}{4\pi} \varepsilon_{abc} \bar{h}^{a}_{\mathbf{k}} \partial_{k_x} \bar{h}^{b}_{\mathbf{k}} \partial_{k_y} \bar{h}^{c}_{\mathbf{k}} ,
\end{eqnarray}
where $\varepsilon_{abc}$ is the total antisymmetric tensor, and $\bar{h}^{a}_{\mathbf{k}}$ are three normalized components of the noninteraction Hamiltonian in the base of the Pauli matrices; i.e., $\hat{h}_{0}(\mathbf{k})=\sum_{a} h_{\mathbf{k}}^{a} \sigma_{a}$, $\sigma_{a}$ are the Pauli matrices, $\bar{h}^{a}_{\mathbf{k}}=h_{\mathbf{k}}^{a}/|h_{\mathbf{k}}|$, $|h_{\mathbf{k}}|^2=\sum_{a} |h_{\mathbf{k}}^{a}|^2$. $\gamma$ is the so-called FDWN. It describes the winding number of the atomic Green's function $G_{\rm at}^{-1}(i\omega)=i\omega + \mu - \Sigma(i\omega)$ on the complex plane. $C_0$ is just the Chern number of the noninteraction Hamiltonian.\cite{Volovik} When the self-energy only depends on frequency, all correlation effects on the Chern number are encoded in the FDWN $\gamma$. One can notice that in the Chern-insulator phase ${\rm Im} G_{\rm at}^{-1}(i\omega)$ crosses the axes $\omega$ one time, whereas in the Mott-insulator phase it never crosses, which leads to $\gamma=1$ in the Chern-insulator phase, and $\gamma=0$ in the Mott insulator.\cite{Xie} Thus, the Mott insulator is topologically trivial. Note that the feature ${\rm Im} \Sigma(i\omega) \sim 1/\omega$ for low frequencies in the Mott-insulator phase is purely a particularity of infinite-dimension systems.\cite{Georges} For finite-dimensional systems, the DMFT is just an approximation and it losses nonlocal correlations. The nonlocal correlations reduce the singularity of the self-energy at low frequencies. We also check whether nonlocal correlations change the Chern number in the insulating phases. We employ an extension of the DMFT, the cellular dynamical mean-field theory (CDMFT),\cite{CDMFT} for calculating the Chern number in both Chern- and Mott-insulator phases. We choose a hexagonal cluster of six sites of the honeycomb lattice and perform the CDMFT calculations.\cite{Liebsch} Within the CDMFT the self-energy matrix is well defined at zero frequency even in the Mott-insulator phase, and we can use Eq. (\ref{chern}) to calculate the Chern number. We obtain $C=1$ in the Chern-insulator phase and $C=0$ in the Mott-insulator phase. This also shows that the topological invariant is robust against the nonlocal correlations. The topological invariant can well be determined within the DMFT, even the DMFT neglects nonlocal correlations. However, the nonlocal correlations slightly change the values of $U_{c1}$ and $U_{c2}$, but  the metallic phase always exists between the two insulator phases. Note that the Hartree-Fock mean-field approximation cannot describe the Mott transition, since it loses dynamical local fluctuations. The DMFT is perhaps the simplest approximation that can capture both the Mott transition and the topological invariant, at least in the proposed Haldane model with local interaction. It is worth checking the applicability of the DMFT to other strong correlation models of topological insulators.

\begin{figure}[t]
\includegraphics[width=0.45\textwidth]{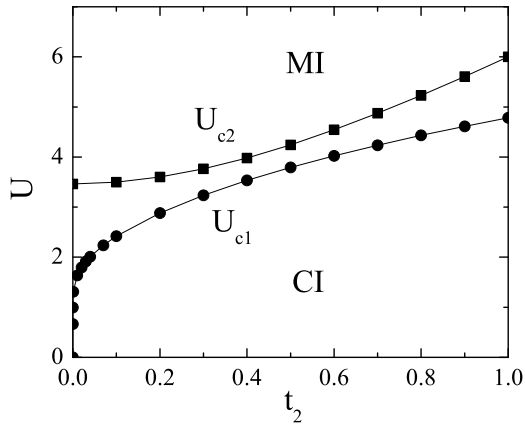}
\caption{The phase diagram of inversion symmetry states at half filling. CI denotes the Chern-insulating phase, and MI denotes the Mott-insulating phase. The pseudogap metallic phase exists between the two phases ($t=1$).} \label{fig3}
\end{figure}

Within the DMFT we can derive explicit equations for determining the critical value $U_{c1}$ and $U_{c2}$. The derivation is based on the linearized DMFT.\cite{Bulla} Due to the particle-hole symmetry at half filling the Green's function is purely imaginary $G(i0^+)=-i\pi \rho(0)$ at the Fermi level. At $U=U_{c1}$, the self-energy is purely real, and $\Delta \Sigma(\omega+i0^+)\equiv \Sigma(\omega+i0^+)-U/2$  becomes negligibly small at $\omega \rightarrow 0$, because
\begin{equation}
G(i0^+) =- \frac{1}{N}\sum\limits_{\mathbf{k}}\frac{\Delta \Sigma (i0^+)}{\left( \Delta \Sigma
(i0^+)\right) ^{2}-t_{2}^{2}f_{2}^{2}(\mathbf{k})-t^{2}|f_{1}(\mathbf{k})|^{2}} \label{zero1}
\end{equation}
approaches to zero. From Eqs. (\ref{gl}) and (\ref{gli}) we obtain
\begin{eqnarray}
G(i0^+) = - \frac{
 \Delta \Sigma (i0^+) U^{2}/4}{\left(\Delta \Sigma (i0^+)\right)
^{2}U^{2}/4-\left( U^{2}/4\right) ^{2}} . \label{zero2}
\end{eqnarray}
At $U=U_{c1}$, $\Delta \Sigma(i0^+)$ vanishes, hence from Eqs. (\ref{zero1})-(\ref{zero2}) we obtain the equation for determining $U_{c1}$
\begin{eqnarray}
\frac{1}{N}\sum\limits_{\mathbf{k}}\frac{1}{
t_{2}^{2}f_{2}^{2}(\mathbf{k})+t^{2}|f_{1}(\mathbf{k})|^{2}}=\frac{4}{U_{c1}^{2}} . \label{uc1}
\end{eqnarray}
At $U=U_{c2}$ the self-energy $\Sigma(i0^+)$ diverges; however the Weiss mean field $\lambda(i0^+)$ vanishes, since from Eqs. (\ref{gl}) and (\ref{gli}), one can show that $\Delta \Sigma(i0^+)=-U^2/\lambda(i0^+)/4$. Together with Eq. (\ref{sc}) we obtain
\begin{equation}
G(i0^+) = \frac{1}{N}\sum\limits_{\mathbf{k}}\frac{\lambda (i0^+)U^{2}/4}{%
\left( U^{2}/4\right) ^{2}-\lambda
^{2}(i0^+)(t_{2}^{2}f_{2}^{2}(\mathbf{k})+t^{2}|f_{1}(\mathbf{k})|^{2})} . \label{zero3}
\end{equation}
On the other hand, from Eq. (\ref{gli}) the local Green's function at zero energy is
\begin{equation}
G(i0^+) = \frac{\lambda (i0^+)}{U^{2}/4-\lambda ^{2}(i0^+)} . \label{zero4}
\end{equation}
At $U=U_{c2}$, $\lambda(i0^+)$ vanishes; hence from Eqs. (\ref{zero3})and(\ref{zero4}) we obtain the equation for determining $U_{c2}$
\begin{equation}
\frac{1}{N}\sum
\limits_{\mathbf{k}}(t_{2}^{2}f_{2}^{2}(\mathbf{k})+t^{2}|f_{1}(\mathbf{k})|^{2})=\frac{U_{c2}^{2}}{4} . \label{uc2}
\end{equation}
In Fig. \ref{fig3} we plot the critical values $U_{c1}$ and $U_{c2}$ as a function of $t_2$. We always obtain $U_{c1}<U_{c2}$. Thus, between the Chern- and the Mott-insulator phases, there is a finite region of the pseudogap metallic phase. There is no direct transition from the topological Chern insulator to the topologically trivial Mott insulator. One can notice that the phase transition from the Chern insulator to the Mott insulator through the pseudogap metallic phase does not change the symmetries of the system. All the phases preserve the inversion symmetry.

So far we have studied the correlation-driven phase transition of the homogeneous phases, where the inversion symmetry is preserved. However these phases are instable to the charge ordering at low temperatures. They are only stable at high temperatures, where the charge long-range order vanishes. At finite temperature, the Hall conductance can be calculated by the Kubo formula\cite{Kawakami}
\begin{eqnarray}
\sigma _{xy} &=&\frac{e^{2}}{\hbar }{\rm Im}\frac{\partial }{\partial
\omega }K(\omega +i0^{+}), \label{kubo1} \\
K(i\omega ) &=&-\frac{T}{N}\sum\limits_{\mathbf{k},i\nu }{\rm Tr}\Big[J_{x}(\mathbf{k}) \hat{G}(\mathbf{k},i\omega +i\nu ) \nonumber \\
&& J_{y}(\mathbf{k})\hat{G}(\mathbf{k},i\nu )\Big],\label{kubo2}
\end{eqnarray}
where $\mathbf{J}(\mathbf{k})$ is the current operator. The Hall conductance is not quantized at finite temperature. However, at zero temperature  the Kubo formula in Eqs. (\ref{kubo1})and(\ref{kubo2}) is reduced to the Chern formula in Eq. (\ref{chern}), and the Hall conductance is a multiple of $e^2/h$.\cite{Kawakami}
At half filling the homogeneous phases do not depend on temperature, and we can consider the Chern invariant as the zero-temperature limit of the Hall conductance calculated by the homogeneous Green's function. The phase diagram which is plotted in Fig. \ref{fig3} is the high-temperature phase transition with the topological invariant determined by the zero-temperature limit of the Hall conductance. It can also be considered as a reference in comparison with the low-temperature charge ordering, as well as in comparison with the paramagnetic phase transition in the correlated $Z_2$ topological insulators.

\section{Charge ordering}
At half filling there is another solution $n_{fA}=1$, $n_{fB}=0$ (or equivalently $n_{fA}=0$, $n_{fB}=1$). Actually, this solution is stable at zero temperature. It is charge ordering, which occurs for any finite value of $U$. The charge ordering breaks the inversion symmetry. This charge ordering at half filling is similar to the one of the Falicov-Kimball model on bipartite lattices.\cite{Mielsch1,Mielsch2,Mielsch3,Freericks,Zlati,Krish,Matveev} From Eq. (\ref{gli}) we obtain
\begin{eqnarray}
G_{A}(z) &=& \frac{1}{G^{-1}_{A}(z)+\Sigma_{A}(z)-U} ,\\
G_{B}(z) &=& \frac{1}{G^{-1}_{B}(z)+\Sigma_{B}(z)} .
\end{eqnarray}
We immediately obtain $\Sigma_{A}(z)=U$ and $\Sigma_{B}(z)=0$. This solution is exactly the Hartree mean-field solution. In infinite dimensions the Hartree mean field becomes exact for the Falicov-Kimball type of the local interaction and at zero temperature. At finite temperatures the self-energies are no longer the Hartree mean-field values, and they actually depend on frequency. With $\Sigma_{A}(z)=U$ and $\Sigma_{B}(z)=0$, the matrix $-\hat{G}^{-1}(i0)$ is exactly the Haldane Hamiltonian with the staggered energy level shift $M=U/2$.\cite{Haldane} There is a topological phase transition from the topological insulator to the topologically trivial insulator at $M_c=3\sqrt{3} t_2$.\cite{Haldane} Thus, for $U<6\sqrt{3} t_2$ the charge-ordered state is topological with the Chern number $C=1$, and for $U>6\sqrt{3} t_2$ it is topologically trivial with $C=0$. At $U=6\sqrt{3} t_2$ the ground state is helical semimetal, where electrons at the corners of the first Brillouin zone are Weyl fermions. We want to emphasis that both the topological insulator and the topologically trivial insulator have the long-range order that is due to electron correlations. Both the phases are charge ordered, but topologically different. Both the phases break the inversion symmetry. The line $U=0$ is the Chern-insulator phase; however it preserves the inversion symmetry. Therefore the topological charge-ordered phase does not adiabatically connect to the Chern insulator at $U=0$. In the topological charge-ordered phase both the topological invariant and the charge long-range order coexist. This contrasts to the case of the same Chern insulator, but with nearest-neighbor interactions, where the charge-ordered phase is topologically trivial.\cite{Galitski1,Galitski2}

\section{Conclusion}
We have studied the electron correlation driven phase transition in the Haldane model with the local Coulomb interaction at half filling. The phase transition depends on the inversion symmetry. With the preservation of the inversion symmetry electron correlations drive the system from the topological Chern insulator, to a pseudogap metal, and then to the topologically trivial Mott insulator. The pseudogap metal is non-Fermi-liquid with renormalized mass and velocity of Weyl fermions. It always exists between the two insulating phases. When the inversion symmetry is broken, electron correlations induce the long-range charge order that opens a gap at the Fermi energy. They drive the system from the topological charge-ordered state to topologically trivial charge-ordered state. The topological invariant and the long-range charge order can coexist due to the effect of electron correlations in the Chern insulator.

The proposed Haldane-Falicov-Kimball model can also be considered as an asymmetry version of the Kane-Mele-Hubbard model, where the time-reversal symmetry is broken. One might expect that when the time-reversal symmetry is broken and the inversion symmetry is preserved, the $Z_2$ topological insulators would exhibit the correlation-driven phase transition from the topological insulator to topologically trivial Mott insulator, and a pseudogap metallic phase would exist between these two insulating phases. When both the time-reversal and the inversion symmetries are broken, there is a possibility of the coexistence of the topological invariant and long-range order. Electron correlations also drive the system from a topological spin-density-wave state to a topologically trivial spin-density-wave state. The time-reversal symmetry in the $Z_2$ topological insulators can be broken by magnetic fields or a mass imbalance of the two spin components. The thin films of Cr-doped (Bi,Sb)$_2$Te$_3$ \cite{Chang,Yu} may relate to this category of topological insulators, where the ferromagnetism plays the role of a magnetic field, and the electron correlations are weak. It is worthwhile to study the possible phase transitions in these thin films by tuning the repulsive Coulomb interaction with and without inversion-symmetry breaking.

\section*{Acknowledgement}

This research is funded by Vietnam National Foundation
for Science and Technology Development (NAFOSTED) under Grant No 103.02-2011.29.

\end{document}